\documentclass[aps,pre,twocolumn,groupedaddress,showpacs]{revtex4-1}

\usepackage{graphics,epsfig}

\begin{document}


\title{Synchronization of Two Diffusively Coupled Chaotic Parametrically Excited nonidentical Pendula}



\author{S. Satpathy}
\email{satya2047@gmail.com}
 \affiliation{Department of Physics \& Astronomy, National Institute of Technology Rourkela, India}
\author{B. Ganguli}%
 \email{biplabg@nitrkl.ac.in}
\affiliation{Department of Physics \& Astronomy, National Institute of Technology Rourkela, India}


\date{\today}

\begin{abstract}
 Auxiliary system approach and various nearest neighbor methods are widely used to detect generalized synchronization in non-identical coupled systems. These methods generally give contradictory results. Therefore one method alone is not sufficient to predict correct result. We show in this report that it is necessary to apply multiple methods together to come to a conclusion. These methods show a signature of generalized synchronization in diffusively coupled non-identical chaotic parametric excited pendula. But we finally find it to be the almost synchronization. It is achieved when the second Lyapunov exponent and both the system's transverse Lyapunov exponents are almost equal. The transition from asynchronous state to almost synchronization is through frequency entrainment as coupling constant is increased. Non-identity of the pendula are realized by mismatch in amplitude of parametric forcing. The frequency entrainment regime does not depend on amplitude mismatch whereas onset of almost synchronization increases with increase in mismatch. The systems 
\end{abstract}

\maketitle


{\bf Generalized synchronization is a common form of synchronization observed in coupled non-identical chaotic oscillators. Yet there exist no single method to confirm this in diffusively coupled systems. Auxiliary system approach and various nearest neighbor methods are widely used to detect generalized synchronization in non-identical coupled systems. Often they give different results. Therefore one method alone is not sufficient to come to the correct result. In the present study we apply auxiliary system approach and various nearest-neighbor methods together with the calculation of Lyapunov exponents to detect the nature of synchronization. For this purpose we take coupled non-identical parametrically excited pendula, non-identity is realized by taking different amplitude of parametric forcing of the two pendula.}

Generalized synchronization (GS) is a fundamental phenomenon in coupled chaotic systems. GS was first introduced for unidirectionally coupled systems \cite{Rulkov1995}. The two systems are in GS if a static functional relation exists between the states of both systems. In \cite{Rulkov1995} a numerical method (mutual false nearest neighbors) is proposed for detecting GS. Since then different techniques have been proposed to detect GS in unidirectionally coupled systems, e.g., other nearest-neighbor methods \cite{Parlitz1996,Parlitz2012,Arnhold1999,Quiroga2000,Schmitz2000,Bhattacharya2003}, the conditional Lyapunov exponent \cite{Pyragas1996}, auxiliary system approach \cite{Abarbanel1996}. Some of these methods are also extended to mutually coupled systems \cite{Boccaletti2000b,Zheng2002,Moskalenko2012}.

The most expected synchronization in non-identical coupled systems is GS. Detection of GS in non-identical systems is not straight forward. It is possible that a single method may not be sufficient to get clear picture of GS. Therefore different methods are required to confirm nature of synchronization.

\begin{figure}[htp]
\includegraphics[scale=1]{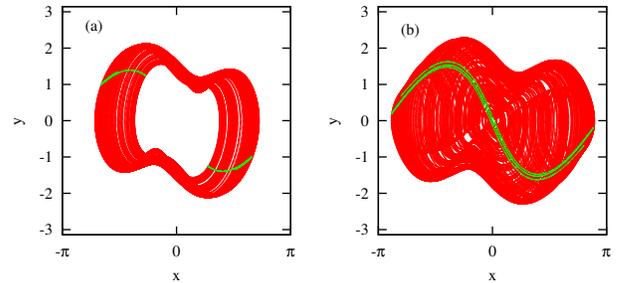}
\caption{Phase portrait for $B = 1.0$, $\omega = 2.0$ and different values of parametric forcing amplitude. (a) $A = 3.25$, (b) $A = 3.55$. The green dots represent the points on the poincar\'e section. }
\label{PPPEP}
\end{figure}

The dynamical equations for the mutually coupled systems in the autonomous form are,
\begin{eqnarray}
\dot{x}_ {1, 2} &=& y_{1, 2}, \nonumber \\ 
\dot{y}_{1, 2} &=& - By_{1, 2} - (1 + A_{1, 2}\cos z)\sin x_{1, 2} \nonumber \\
&&+  k(x_{2, 1} - x_{1, 2}), \\ 
\dot{z} &=& \omega, \nonumber
\label{CNIPEP}
\end{eqnarray}

\noindent where the subscripts $1$ and $2$ refer to pendulum $1$ and $2$, respectively, and $k$ is the coupling constant.  $B$ , $A$, and $\omega$ are the scaled forms of damping coefficient, amplitude of parametric forcing, and frequency of parametric forcing respectively. $B= 1.0$, $A_1 = 3.25$, $A_2 = 3.55$ and $\omega = 2.0$, are the parameters chosen to produce chaotic behavior. The phase trajectories of the two systems for $k=0$ are shown in Fig. \ref{PPPEP} \cite{Satpathy2017}.

\begin{figure}[htp]
\includegraphics[scale=1]{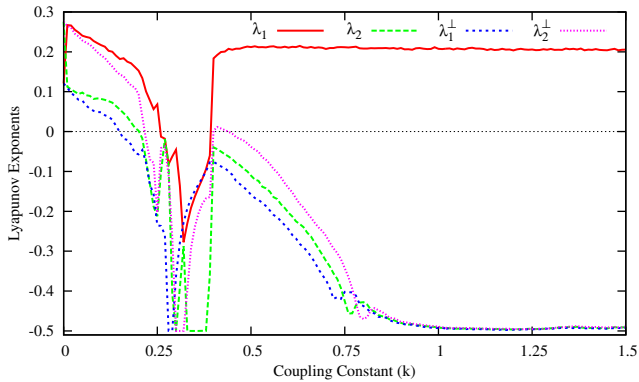}
\caption{The two largest Lyapunov exponents ($\lambda_1$, $\lambda_2$) and the system's transverse Lyapunov exponents ($\lambda_1^\perp$, $\lambda_2^\perp$) vs coupling strength $k$}
\label{LNIB}
\end{figure}

\begin{figure}[htp]
\includegraphics[scale=1]{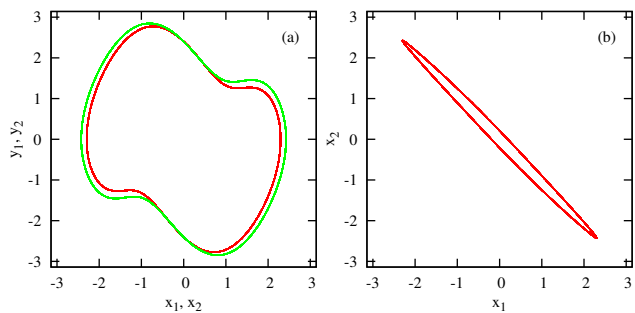}
\caption{(a) Projection of the phase trajectories of the two systems in $(x,y)$ plane. (b) Projection of the phase trajectories in $(x_1, x_2)$ plane.}
\label{PFE}
\end{figure}

We calculate Lyapunov \cite{Wolf1985} and system's transverse Lyapunov exponents \cite{Acharyya2011} as a function of coupling constant to study the stability of synchronized state. The Lyapunov spectrum have two positive, two negative, and one zero exponents at $k = 0$. The zero exponent is due to non-autonomous nature of the original system and it is insensitive to the coupling. The two largest Lyapunov exponents ($\lambda_1$, $\lambda_2$), and the largest system's transverse Lyapunov exponents ($\lambda_1^\perp$, $\lambda_2^\perp$)  are plotted as a function of coupling constant, shown in Fig. \ref{LNIB}. As the coupling constant $k$ increases, $\lambda_2$ becomes negative at $k = 0.21$. This indicates a bifurcation from hyper-chaotic to chaotic state. $\lambda_1^\perp$ is negative for $k \geq 0.15$ and $\lambda_2^\perp$ is negative for $0.22 \leq k \leq 0.39$ and also for $k \geq 0.44$. $\lambda_1$  is negative for $0.26 \leq k \leq 0.39$. This means pendula have periodic motions in this range of coupling. The closed curve between $x_1$ and $x_2$ in Fig. \ref{PFE} shows frequency entrainment in this regime. $\lambda_1^\perp$ and $\lambda_2^\perp$ are both simultaneously negative for $k \geq 0.44$. Therefore synchronization occurs in this range of coupling constant.

We first use auxiliary system approach for the mutually coupled system to detect GS as discussed in \cite{Abarbanel1996,Zheng2002}. Due to the bidirectional interaction, one should introduce two auxiliary systems $\mathbf x^\prime$ and $\mathbf y^\prime$ , which, respectively, are identical to $\mathbf x \equiv(x_1, y_1, z) $ and $\mathbf y \equiv(x_2, y_2, z)$, i.e., let $\mathbf x$ drive $\mathbf y^\prime$ and $\mathbf y$ drive $\mathbf x^\prime$. $\mathbf x^\prime$ and $\mathbf y^\prime$ possess the same parameters as $\mathbf x$ and $\mathbf y$, but evolve from different initial conditions. Thus the vector fields in the phase spaces of $\mathbf x$ (or $\mathbf y$) and $\mathbf x^\prime$ (or $\mathbf y^\prime$) are identical and they can evolve on identical attractors. With increasing the coupling strength, one may expect both $\mathbf x-\mathbf x^\prime$ and $\mathbf y-\mathbf y^\prime$ tends to zero after the initial transients. To find GS we consider the average distances of auxiliary systems from their respective CS manifold and its maximum values. The average distances of auxiliary systems from their respective CS manifold are defined by
\begin{eqnarray}
D_{1} &=& \lim_{t \to \infty}\frac{1}{T - T_0}\int_{T_0}^T \mid {\mathbf x}(t) - {\mathbf x^\prime}(t) \mid dt, \\
D_{2} &=& \lim_{t \to \infty}\frac{1}{T - T_0}\int_{T_0}^T \mid {\mathbf y}(t) - {\mathbf y^\prime}(t) \mid dt. 
\end{eqnarray}
where $T$ is the time of the calculation, and $T_0$ is the transient time.

The equations of the auxiliary systems are
\begin{eqnarray}
\dot{x}^\prime_ {1, 2} &=& y^\prime_{1, 2}, \nonumber \\ 
\dot{y}^\prime_{1, 2} &=& - By^\prime_{1, 2} - (1 + A_{1, 2}\cos z)\sin x^\prime_{1, 2}  \nonumber \\
&&+  k(x_{2, 1} - x^\prime_{1, 2}), \\ 
\dot{z} &=& \omega, \nonumber
\end{eqnarray}

\begin{figure}[htp]
\includegraphics[scale=1]{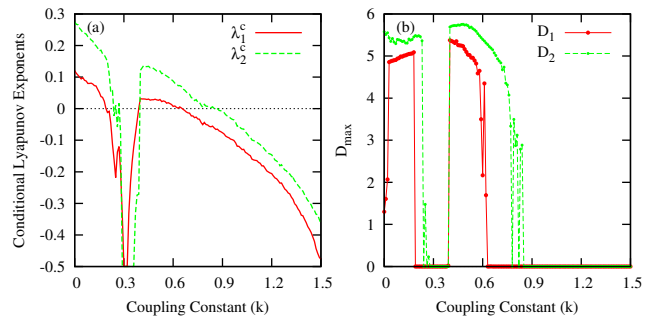}
\caption{(a) The systems are synchronized when the largest conditional Lyapunov exponents of the two systems are negative.  (b) The maximum of averaged distances $D$ approach zero indicating the occurrence of synchronization.}
\label{ASA}
\end{figure}

The largest conditional Lyapunov exponents ($\lambda_1^c$ and $\lambda_2^c$) of the two systems and the maximum values of the average distances ($D_1$ and $D_2$) from the CS manifold are plotted in Fig.  \ref{ASA}. Both the distances are zero and the largest conditional Lyapunov exponents are negative in the frequency entrainment regime. The figure shows two thresholds of the onset of GS. The first threshold is at $k = 0.66$, where both $D_1 \to 0$ and $\lambda_1^c < 0$ are simultaneously satisfied. This indicates the onset of partial GS. At the first threshold the drive $2$ synchronizes the auxiliary system $1^\prime$ to the system $1$. Therefore GS is achieved first in one of the the coupling direction, i.e., the system $1$ is slaved by the system $2$ for mutual coupling. At the second threshold, $k = 0.86$, both $D_2 \to 0$ and $\lambda_2^c < 0$ are simultaneously satisfied. This indicates the onset of global GS. We find that the system with $A = 3.25$ and $\lambda  = 0.12384$ is synchronized first, than the other one with $A = 3.55$ and $\lambda = 0.27713$. It has been shown that auxiliary system approach may not correctly give GS state in mutually coupled system  \cite{Moskalenko2013}. Further calculations are necessary for verification.

The second method is based on nearest neighbor described in \cite{Parlitz1996,Parlitz2012,Moskalenko2012}. Let $U$ and $V$ be the embedding space of ${\mathbf x}$ and ${\mathbf y}$ respectively. If there exist any functional relation between the reconstructed states ${\mathbf u}^n$ and ${\mathbf v}^n$, then any neighboring states of ${\mathbf u}^n$ are mapped to neighbors of ${\mathbf v}^n$ and vice versa. As a numerical indicator of the existence of a functional relationship between the interacting systems, we have selected the nearest neighbor ${\mathbf u}^{nn}$ of ${\mathbf u}^n$ for $n = 1,.\,.\,.\, ,N$ and have computed the average distance of the corresponding image points ${\mathbf v}^n$ and ${\mathbf v}^{nn}$. This mean distance between images of nearest neighbors is normalized by the average distance $\delta_\mathbf y$ of randomly chosen states of the second system, i.e.,
\begin{equation}
d_{\mathbf{xy}} = \frac{1}{N\delta_\mathbf{y}}\sum_{n=1}^N \parallel{\mathbf v}^n-{\mathbf v}^{nn}\parallel.
\label{MD1}
\end{equation}

Analogously, we look in the opposite direction whether nearest neighbors in $\mathbf y$-space are mapped to nearest neighbors in $\mathbf x$-space
\begin{equation}
d_{\mathbf {yx}} = \frac{1}{N\delta_\mathbf x}\sum_{n=1}^N \parallel{\mathbf u}^n-{\mathbf u}^{nn}\parallel.
\label{MD2}
\end{equation}

This characteristic allows us to reveal the qualitative changes in the synchonous or asynchonous behavior of the coupled systems. When the coupling between systems is very small and oscillators show the asynchronous dynamics, the value of this measure, $d \sim 1$. $d$ tends to be zero deep inside the generalized synchronization region due to the presence of the functional relation between states of the interacting systems. Unfortunately, the nearest neighbor method does not allow us to detect precisely the boundary points of the GS regime, but it allows us to confirm the presence of GS.

\begin{figure}[htp]
\includegraphics[scale=1]{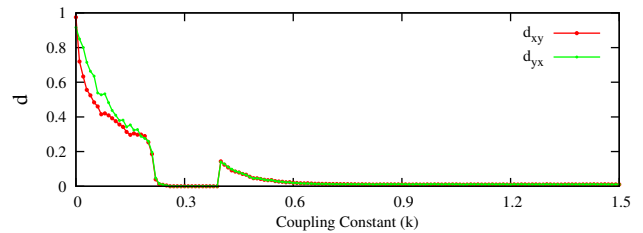}
\caption{The quantitative measure $d$ vs the coupling parameter strength $k$.}
\label{ContinuityB}
\end{figure}

We reconstructed the attractors by sampling the data $x_1$ and $x_2$. Sampled data are chosen at time step  $45\Delta t$. $\Delta t$ is the time step of integration. The time delays calculated from mutual information are taken as $4$. The embedding dimension for system $1$ and for $k > 0.15$ is $4$ whereas for $k \leq 0.15$ is $5$. Similarly the embedding dimension for system $2$ and for $k > 0.14$ is $4$ whereas for $k \leq 0.14$ is $5$. With this data, we calculate the average distance $d_\mathbf{xy}$ and $d_\mathbf{yx}$. The variation of $d_\mathbf{xy}$ and $d_\mathbf{yx}$ as a function of coupling constant is shown in Fig. \ref{ContinuityB}. In the frequency entrainment regime both $d_\mathbf{xy}$ and $d_\mathbf{yx}$ tend to zero. At further higher coupling also both $d_\mathbf{xy}$ and $d_\mathbf{yx}$  tend to zero. This confirms GS though it can not predict the boundary of onset of GS.

The third method is the variation of false nearest neighbor method \cite{Arnhold1999,Quiroga2000,Schmitz2000,Bhattacharya2003}. Here the difference is to consider $k$ nearest neighbors of the first system ${\bf x}$ and of the second system ${\bf y}$. Let $X$ and $Y$ be the time series of the first and second systems in the embedding space respectively. Let $n_{NN1}(j) (j = 1, ..., k)$ and $n_{NN2}(j) (j = 1, ..., k)$ be the time index of the $k$ nearest neighbor of  ${\bf x}_n$ and ${\bf y}_n$ respectively. For each ${\bf x}_n$, the squared mean Euclidean distance to its $k$ closest neighbors is defined as

\begin{equation}
R_n^{(k)}(X) = \frac{1}{k} \sum_{j = 1}^k ({\bf x}_n - {\bf x}_{n_{NN1}(j)})^2,
\end{equation}
while the conditional mean squared Euclidean distance, conditioned on the closest neighbor times in the time series Y, is
\begin{equation}
R_n^{(k)}(X|Y) = \frac{1}{k} \sum_{j = 1}^k ({\bf x}_n - {\bf x}_{n_{NN2}(j)})^2.
\end{equation}

With these quantities, a global interdependence measures can be defined as 
\begin{equation}
\label{SXY}
S^{(k)}(X|Y) = \frac{1}{N} \sum_{n = 1}^N\frac{R_n^{(k)}(X)}{R_n^{(k)}(X|Y)}.
\end{equation}

If $S^{(k)}(X|Y) \ll 1$ then  $X$ and $Y$ are independent or unsynchronized and $S^{(k)}(X|Y) \to 1$ indicates the occurance of  GS. The opposite dependence $S^{(k)}(Y|X)$ is defined in complete analogy. Both the dependences are in general not equal. $S^{(k)}(X|Y) > S^{(k)}(Y|X)$ implies that  $X$ depends more on $Y$ than vice versa, i.e.  $Y$ is more active than $X$.

In Eq. \ref{SXY} we compare the Y-conditioned mean squared distances to the mean squared nearest neighbor distances. Instead of this, we could have compared the former to the mean squared distances to random points,
\begin{equation}
R_n(X) = \frac{1}{N-1} \sum_{j \neq n}({\bf x}_n - {\bf x}_j)^2.
\end{equation}

The geometrical average, in analogy to the Eq. \ref{SXY} is defined as
\begin{equation}
H^{(k)}(X|Y) = \frac{1}{N}\sum_{n = 1}^N \frac{R_n(X)}{R_n^{(k)}(X|Y)}.
\end{equation}

It is zero if $X$ and $Y$ are completely independent, while it is positive if nearness in $Y$ also implies nearness in $X$. It would be negative if close pairs in $Y$ corresponded mainly to distant pairs in $X$. This is very unlikely but not impossible. $H^{(k)}(Y|X)$ is defined in complete analogy.

\begin{figure}[htp]
\includegraphics[scale=1]{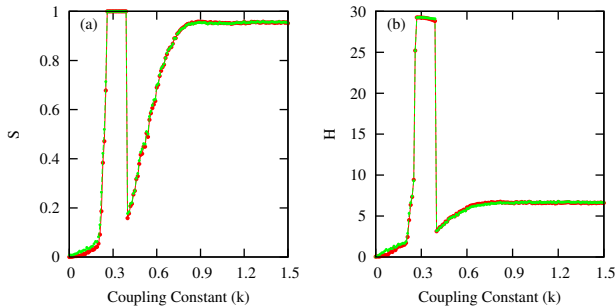}
\caption{Red: $S(X|Y)$ and $H(X|Y)$; Green: $S(Y|X)$ and $H(Y|X)$. Flat regions corresponds to synchronization regime.}
\label{SHEB} 
\end{figure}

The similarity indexes $S$ and $H$ are calculated  from the already reconstructed attractors. $S(X|Y)$, $S(Y|X)$, $H(X|Y)$ and $H(Y|X)$ for $10$ nearest neighbors are plotted as a function of coupling constant, shown in Fig. \ref{SHEB}. $S$ remains near to $1$ and $H$ remains constant in the frequency entrainment regime. They remain constant at some higher range of coupling constant which signifies GS.

The fourth method is mutual false nearest neighbor (MFNN) parameter \cite{Rulkov1995,Boccaletti2000b}. Consider three embedding spaces $S_1$,  $S_2$ and $S_3$.  $S_1$ is the embedding space of the first system at fixed embedding dimension $d_1$, 
$S_2$ is the embedding space of the second system at variable embedding dimension $d_2$ and $S_3$  is the embedding space of the second system at the fixed embedding dimension $d_1$. 

We then chooses randomly $n$ state vectors ${\mathbf y}(n)$ at time $t_n$ in $S_2$ and consider their images ${\mathbf x}(n)$ and ${\mathbf x^\prime}(n)$ in  $S_1$ and $S_3$ respectively. Let the time index of the nearest neighbor in $S_1$, $S_2$, $S_3$ be  $n_{NN1}$, $n_{NN2}$, $n_{NN1^\prime}$ respectively. The MFNN parameter is then defined as
\begin{equation}
r = \left\langle \frac{ \vert{\mathbf x}(n) - {\mathbf x}(n_{NN2})\vert^2  }{ \vert{\mathbf x}(n) - {\mathbf x}(n_{NN1})\vert^2 }\frac{ \vert{\mathbf x}^\prime(n) - {\mathbf x}^\prime(n_{NN1^\prime})\vert^2 }{ \vert{\mathbf x}^\prime(n) - {\mathbf x}^\prime(n_{NN2})\vert^2 }\right\rangle_n,
\end{equation}
where $\langle ...\rangle_n$ denotes the averaging process over $n$. $r \equiv 1$ for the systems showing GS, whereas $r \neq 1$ for unsynchronized systems. 

\begin{figure}[htp]
\centering
\includegraphics[scale=1.0]{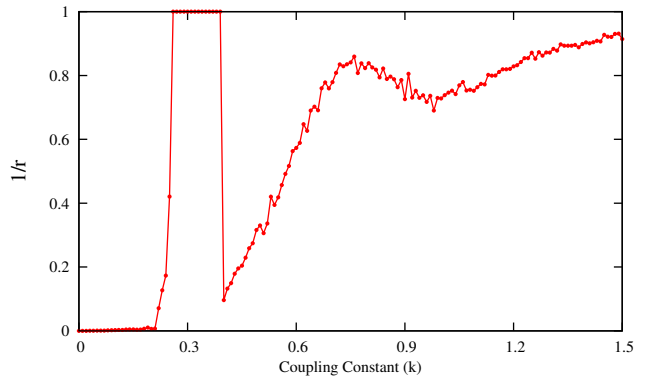}
\caption{$\bar{r}$ as a function of coupling strength $k$.}
\label{MFNNNB}
\end{figure}

The embedding dimension of the first system, $d_1$,  is taken as before, i.e., $4$ for $k > 0.15$, $5$ for $k \leq 0.15$, which is kept fixed through out the calculation. MFNN parameter is then calculated by varying the dimension of the second system, $d_2$. When the attractor is completely unfolded the MFNN parameter approaches $1$. The dimension of the second system is taken $10$ for further calculation which is larger than the dimension required to unfold the attractor. The MFNN parameter as a function of coupling constant is calculated. The inverse of the MFNN parameter ($r^{-1}$) is shown in Fig. \ref{MFNNNB}. $r^{-1} \sim 1$ in the frequency entrainment regime. For higher coupling range it approaches to $1$, but not steady. This means GS is unstable with respect to coupling constant.

\begin{figure}[htp]
\includegraphics[scale=1.0]{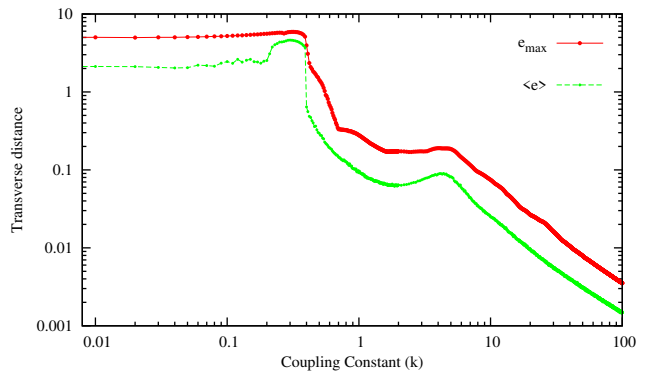}
\caption{The mean synchronization error $\langle e \rangle$ and its maximum value  $e_{max}$ as a function  of coupling strength $k$.}
\label{ErrorsB}
\end{figure}

\begin{figure}[htp]
\includegraphics[scale=1.0]{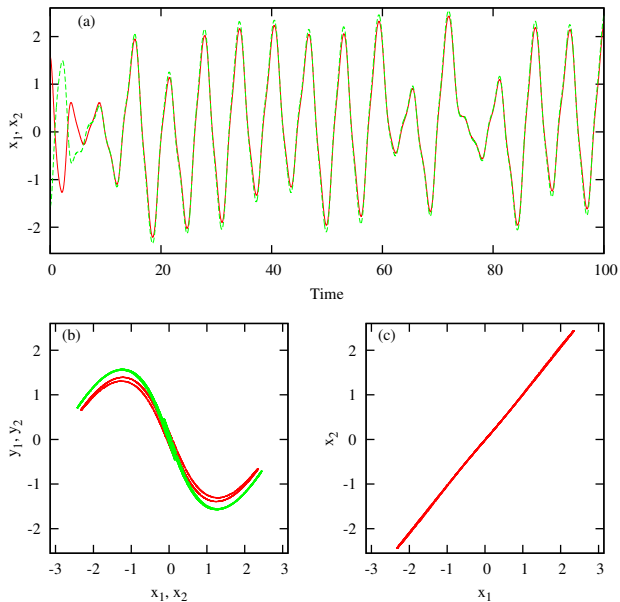}
\caption{(a) Time series of the two systems. (b) The points on the poincar\'e map in the plane $(x_1, y_1)$ (red), $(x_2, y_2)$ (green) (c) in the plane $(x_1, x_2)$.}
\label{TIMESB}
\end{figure}

The results obtained by various methods are somewhat contradictory. This requires further calculation for confirmation of the synchronized state. For this purpose we calculate the mean synchronization error $\langle e \rangle$ and its maximum value  $e_{max}$ \cite{Boccaletti2002}, shown in Fig. \ref{ErrorsB} as a function of coupling constant, which is given by
\begin{equation}
\langle e \rangle  = \lim_{T \to \infty}\frac{1}{T - T_0}\int_{T0}^T \mid {\mathbf x}(t) - {\mathbf y}(t) \mid dt,
\end{equation}
It is clear that these values decreases with increase in coupling strength. This indicates the synchronized state is closed to CS known as almost synchronization (AS) \cite{Afraimovich1986,Femat1999}.  The time series plot, Fig. \ref{TIMESB}(a), of the variables $x_1$ and $x_2$  shows complete phase matching but the amplitudes are slightly different. A slight difference in amplitude is reflected in Fig. \ref{TIMESB}(b \& c).

\begin{figure}[htp]
\centering
\includegraphics[scale=0.75]{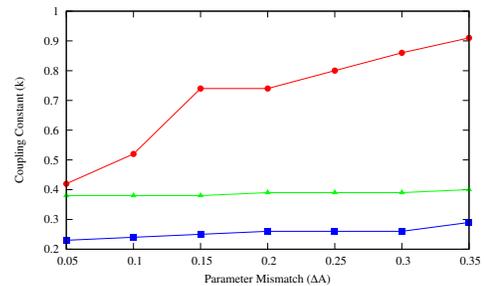}
\caption{Onset of AS (red line with circle);  frequency entrainment region between blue line (squares) and green line (triangles).}
\label{BAK}
\end{figure}

Among all the methods, it is only auxiliary system approach gives boundary of synchronized states. We take the boundary of the global GS found earlier as the onset of AS. The calculations are repeated for different set of systems where the amplitude of parametric forcing of the second system is varied. We summarize the results in Fig. \ref{BAK}. The figure shows the coupling range of frequency entrainment remains almost same when the mismatch of amplitude of parametric forcing is varied. But the coupling constant for onset of AS increases with increase in mismatch of amplitude of parametric forcing.

In summary we studied GS in a coupled parametrically excited chaotic non-identical pendula. Auxiliary system approach and methods based on different variations of false nearest neighbor predict GS whereas mutual false nearest neighbor parameter shows GS state to be unstable with respect to coupling constant. This contradictory result shows the GS may be of special kind. Calculation of synchronization error confirms the GS sate to be AS. Since it is only auxiliary system approach provides boundary, therefore boundary of AS is taken to be same as this. We conclude that when GS is of special kind like AS, methods of detecting GS may give contradictory result.

Frequency entrainment of order 1:1 is observed when the systems show periodic behavior. The system achieves AS when the second Lyapunov exponent and both the system's transverse Lyapunov exponents are almost equal. When the mismatch in amplitude of parametric forcing ($\Delta A$) of the two oscillators is increased, the range of coupling constants for frequency entrainment remains almost the same. The coupling constant for onset of AS increases with increase in $\Delta A$.

\bibliography{Bibliography}

\end{document}